\documentclass[showpacs,amssymb,aps]{revtex4}
\usepackage{graphicx}
\begin{document}
\title{Coulomb screening in graphene with topological defects}

\author{Baishali Chakraborty}
\email{baishali.chakraborty@saha.ac.in}

\author{Kumar S. Gupta}
\email{kumars.gupta@saha.ac.in}
\affiliation{Theory Division, Saha Institute of Nuclear Physics, 1/AF Bidhannagar, Calcutta 700064, India}
\author{Siddhartha Sen}
\email{siddhartha.sen@tcd.ie}
\affiliation{CRANN, Trinity College Dublin, Dublin 2, Ireland}
\date{\today}

\begin{abstract}
We analyze the screening of an external Coulomb charge in gapless graphene cone, 
which is taken as a prototype of a topological defect. In the subcritical regime, 
the induced charge is calculated using both the Green's function and the Friedel 
sum rule. The dependence of the polarization charge on the Coulomb strength obtained 
from the Green's function clearly shows the effect of the conical defect and indicates 
that the critical charge itself depends on the sample topology. Similar analysis using 
the Friedel sum rule indicates that the two results agree for low values of the Coulomb 
charge but differ for the higher strengths, especially in the presence of the conical 
defect. For a given subcritical charge, the transport cross-section has a higher value in the presence of the conical defect. 
In the supercritical regime we show that the coefficient of the power law tail of polarization charge density can be expressed as a 
summation of functions which vary log periodically  with the distance from the Coulomb 
impurity. The period of variation depends on the conical defect. In the presence of the conical defect, 
the Fano resonances begin to appear in the transport cross-section for a lower value of the Coulomb charge. 
For both sub and supercritical regime we derive the dependence of LDOS on the conical defect. 
The effects of generalized boundary condition on the physical observables are also discussed.

\end{abstract}
\maketitle

\section{Introduction}
Coulomb screening in graphene \cite{kats3,ando,levi1,levi2} provides a glimpse of the strong nonperturbative 
QED effects \cite{QED,levi1,levi2,kats1,us1,rmp4} and gives information about the measurable transport 
properties \cite{nomura,hwang,chen,rmp2,rmp3,radchenko}. The phenomenon of the 
Coulomb screening can be analyzed in several ways. One method is to use the Green's function technique \cite{green3,martin,brown,mil1,mil3,mil2}, 
which was originally developed using the solutions of the corresponding eigenvalue equation \cite{green3,martin}. Subsequently a more powerful operator 
formalism for the Green's function was introduced in QED \cite{brown,mil1,mil3}, 
which has been adapted to the case of planar graphene \cite{mil2}. Another approach for studying Coulomb screening is provided 
by Friedel sum rule \cite{friedel,mahan}, which uses the spectral data in the scattering sector. In this approach, the induced 
charge is related to the scattering phase shift. This technique has also been applied to a variety of fermionic systems \cite{moroz1,moroz2,lin1,lin2} 
including planar graphene \cite{lin3,levi1}. 

If the external Coulomb charge in graphene exceeds a certain critical value, the system exhibits quantum instability. 
This is characterized by a rapid oscillatory behaviour of the 
wavefunction near the location of the charge \cite{levi1,levi2}. In addition, 
quasi-bound states appear in the spectrum for gapless graphene \cite{levi2,kats1,us1}, 
for which there is recent experimental evidence \cite{levi3}. 
These phenomena are manifestations of strong nonperturbative QED 
effects. 

Topological defects provide another source of nonperturbative quantum effects in graphene \cite{voz2,crespi1,crespi2,osi1,
sitenko,voz3,furtado,voz5,stone2,critical1}. 
A graphene cone provides a simple prototype of a topological defect, which can be obtained by cutting a sector of 
planar graphene and gluing the two edges of the removed sector. This conical defect generates nontrivial holonomies 
of the quasiparticle wavefunction \cite{voz2,crespi1,crespi2,sitenko,voz3,furtado,voz5,critical1}, 
which can be modelled through the introduction of a fictitious magnetic flux tube with a suitable vector 
potential. The problem of the Coulomb charge in a graphene cone can therefore be related 
to an equivalent problem of the Coulomb charge in planar graphene with a suitably chosen magnetic flux tube. 

The main purpose of this paper is to show how the conical defect affects the Coulomb screening and related 
physical observables in gapless graphene. In the subcritical regime of the external charge, the screening 
is analyzed using both the Green's function and the Friedel sum rule. 
We follow the operator technique \cite{mil1,mil3,mil2} to calculate the Green's function from which the induced charge density is obtained. 
Our calculation clearly shows the effect of the conical defect on Coulomb screening. In particular, 
a plot of the induced charge against the Coulomb strength indicates how the conical defect modifies 
the value of the critical charge. We also calculate the induced charge using the Friedel sum rule \cite{levi1}. 
For the case of planar graphene, these two approaches agree for low values of the external 
charge and starts to show divergence as the critical value of the charge is approached. 
In the presence of the conical defect, the difference shows up at much lower value of the external 
charge. This could be indicative of certain subtleties associated with the Friedel sum rule in 
the presence of singularities \cite{graphs}. For both the Green's function and Friedel sum rule 
approach, our calculations with the conical defect have the correct planar limit. For a given 
value of the subcritical charge, the transport cross-section has higher value in the presence 
of the topological defect. Similarly, the LDOS is also affected by the topological defect.

In the supercritical regime, the induced charge for planar graphene has already been obtained in the literature using both the Green's 
function \cite{nishida} and the Friedel sum rule \cite{levi1}. Here we show how the topological defect affects 
the scattering phase shifts in the supercritical regime. This leads to the corresponding effect on the induced 
charge via the Friedel sum rule. The polarization charge density in the supercritical regime exhibits a power 
law tail. The coefficient of the power law tail can be written as a sum of functions 
which vary log periodically with the distance from the Coulomb impurity \cite{nishida}. We show 
that this period is affected by the presence of the conical defect. In the presence of the topological 
defect, the Fano resonances \cite{levi1,levi3} appear for a lower value of the charge. This indicates 
that the conical topology affects the value of the critical charge. We also calculate the LDOS and 
find its dependence on the conical defect. These results could be of interest in the context of 
graphene based electronic devices \cite{novo1,naturenano}. In the supercritical regime also, our results have the correct planar limit.

For certain parameter ranges in the subcritical regime, the gapless graphene system admits generalized 
boundary conditions \cite{critical2,ksg1,critical3}. When a Coulomb charge is embedded in a 
graphene sheet, it can induce various short range interactions. 
Effect of these interactions cannot be directly incorporated in  the Dirac equation 
which is valid only in the long wavelength limit. The generalized boundary conditions encode an average effect of such short range interactions. 
The physical observables of the system depend on the choice of these generalized boundary 
conditions. Here we have shown how such generalized boundary conditions affect the LDOS and transport cross-section in the presence of the topological defect.

This paper is organized as follows. In Section 2 we calculate the induced charge
using the Green's function in a subcritical Coulomb field in graphene cone.
Next we analyze the subcritical region using the Friedel sum rule
and show how the physical quantities depend explicitly on the sample topology.
We discuss the effect of topological defect on the electrical conductivity of the system.
We compare our results obtained using Green's function technique and Friedel sum rule. 
In Section 3 the analysis of the corresponding spectrum is done in the supercritical region.
This is followed by a discussion of the effect of generalized boundary conditions in the subcritical regime in Section 4. 
We conclude this paper with a summary in Section 5.

\section{Screening of subcritical Coulomb charge in graphene with conical defect}
\subsection{Induced charge from Green's function}

Our system consists of an external Coulomb charge impurity at the apex of a gapless graphene cone. 
In order to set up the formalism, we first consider planar graphene. For technical reasons \cite{mil3,mil2} 
it is useful to introduce a mass $\mathcal{M}$ for the quasiparticles in graphene. This serves as an infrared 
cutoff which is removed at the end of the calculations. 

The Dirac equation for a planar graphene in presence of a Coulomb charge is given by
\begin{eqnarray}
\label{g}
 \left[-i\sigma_1 \partial_x - i\sigma_2 \partial_y-\frac{\alpha}{r} + \mathcal{M} \sigma_3\right]\Psi = E\Psi,
\end{eqnarray}
where the wavefunction $\Psi$ has two components corresponding to the two different sublattices.
The Pauli matrices $\sigma_{1,2,3}$ act on the sublattice 
indices and $\alpha$ is the Coulomb potential strength. A positive value of $\alpha$ corresponds to 
a positive external charge. The radial coordinate in the $x-y$ plane is denoted by $r$. 
The Dirac equation can be written in the form
\begin{eqnarray}
\label{g5}
 [\hat{\mathcal{H}}_0-\mathcal{M}]\Psi=0, ~~~\mbox{where}~~~\hat{\mathcal{H}}_0=\sigma_3 \left[E-\left(-i\sigma_1 \partial_x - i\sigma_2 \partial_y-\frac{\alpha}{r}\right)\right].
\end{eqnarray}
The Green's function $G(\vec{r},\vec{r_0};E)$ corresponding to Eq.(\ref{g5}) satisfies the equation 
\begin{eqnarray}
\label{gf1}
[\hat{\mathcal{H}}_0-\mathcal{M}]G(\vec{r},\vec{r_0};E)=\delta(\vec{r}-\vec{r_0})
\end{eqnarray}
and the induced charge density $\rho_I(\vec{r})$ in graphene is given by
\begin{eqnarray}
 \label{g16}
\rho_I(\vec{r}) = -4ie\int_C \frac{dE}{2\pi}\mbox{Tr}\{G(\vec{r},\vec{r};E)\},
\end{eqnarray}
where the factor $4$ appears due to spin and valley degeneracies in graphene, $e$ is the electronic charge,
$G(\vec{r},\vec{r};E)$ denotes the Green's function and $C$ denotes the contour of integration. 
The Green's function defined in Eq.(\ref{gf1}) can be written as 
\begin{eqnarray}
 \label{g8.1}
G(\vec{r},\vec{r_0};E)=-i[\hat{\mathcal{H}}_0+\mathcal{M}][-ir\{\hat{\mathcal{H}}_0^2-\mathcal{M}^2\}]^{-1}\sqrt{\frac{r}{r_0}}\delta(r-r_0)\delta(\theta-\theta_0),
\end{eqnarray}
where $\delta(\vec{r}-\vec{r_0})=\frac{1}{\sqrt{rr_0}}\delta(r-r_0)\delta(\theta-\theta_0)$ in polar coordinates.
Using the Laplace transformation
\begin{eqnarray}
 \label{g9}
[-ir\{\hat{\mathcal{H}}_0^2-\mathcal{M}^2\}]^{-1}=\int_{0}^{\infty}ds e^{irs\{\hat{\mathcal{H}}_0^2-\mathcal{M}^2\}},
\end{eqnarray}
the Green's function can be written as
\begin{eqnarray}
\label{g9.1}
G(\vec{r},\vec{r_0};E)=-i[\hat{\mathcal{H}}_0+\mathcal{M}]\int_{0}^{\infty}ds e^{irs\{\hat{\mathcal{H}}_0^2-\mathcal{M}^2\}}\sqrt{\frac{r}{r_0}}\delta(r-r_0)\delta(\theta-\theta_0). 
\end{eqnarray}
\begin{figure}
[ht] 
\centering
\includegraphics[bb= 320 54 12 210]{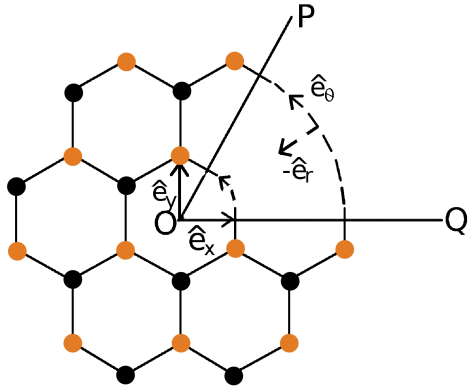}
\caption{Formation of a graphene cone from a plane sheet of graphene by removing 
         the sector $POQ$ and identifying the edges $OP$ and $OQ$.}
\label{fig:1}
\end{figure}

We now apply the above formalism to graphene with a conical defect. As shown in Fig.(1), such a 
defect is introduced in planar graphene by removing $n$ sectors subtending an angle $\frac{2n\pi}{6}$ 
and then identifying the edges, where $n$ can have values $1,2,3,4,5$. The value $n=0$ denotes planar 
graphene. As a result of this identification, the frame $\{\hat{e}_x,\hat{e}_y\}$, 
shown in Fig.(\ref{fig:1}), becomes discontinuous across 
the joining line. To take care of this discontinuity we choose a new set of rotated frames \cite{crespi1,crespi2,critical1} 
$\hat{e}_{x^{\prime}}=\hat{e}_{\theta}$ and $\hat{e}_{y^{\prime}}=-\hat{e}_{r}$. 
The wavefunction associated with a quasiparticle in graphene picks up a holonomy when the quasiparticle encircles the conical defect.
This is a topological effect, which can be modelled by considering a graphene plane with a fictitious flux tube perpendicular to the 
planar surface. The vector potential ${\vec{A}}$ associated with this fictitious flux tube is so chosen that when a quasiparticle 
encircles the flux tube, the wavefunction picks up exactly the same holonomy as in the case of the graphene cone. The resulting Dirac equation is given by
\begin{eqnarray}
 \label{g1}
\bigg [ \vec{\sigma}.\hat{p}^{\prime}-e\sigma_{\theta}A_{\theta}+\mathcal{M}\sigma_3-\frac{\alpha}{r} \bigg ] \Psi = E \Psi
~~~~\mbox{where}~~~~~ A_\theta=\frac{1}{er}\left[\frac{\pm\frac{n}{4}}{(1-\frac{n}{6})}+\frac{\sigma_3}{2}\right].
\end{eqnarray}
Here $\hat{p}^{\prime}$ represents the momentum in the primed coordinate system with \cite{crespi1,crespi2,critical1} 
\begin{eqnarray}
\label{g3}
 \vec{\sigma}.\hat{p}^{\prime}=i\sigma_2\partial_r - \frac{i\sigma_1}{r(1-\frac{n}{6})}\partial_\theta
\end{eqnarray}
and
\begin{eqnarray}
\label{g4}
\sigma_{\theta}A_{\theta}=\frac{1}{er}\left[\frac{\pm\frac{n}{4}\sigma_1}{(1-\frac{n}{6})}-\frac{i\sigma_2}{2}\right]. 
\end{eqnarray}
Using Eq.(\ref{g3}) and Eq.(\ref{g4}), the Dirac equation in the presence of the conical defect can be written as 
\begin{eqnarray}
 \label{g7}
[\hat{\mathcal{H}}-\mathcal{M}]\Psi=0, ~~~\mbox{where}~~~
\hat{\mathcal{H}}= \sigma_3\left(E+\frac{\alpha}{r}\right)-\frac{i\sigma_2}{r(1-\frac{n}{6})}\left(-i\partial_\theta \pm \frac{n}{4}\right) - \sigma_1 \left(\partial_r + \frac{1}{2r}\right).
\end{eqnarray}
From Eq.(\ref{g7}) we get 
\begin{eqnarray}
 \label{g10}
\hat{\mathcal{H}}^2-\mathcal{M}^2 = \left(\partial^{2}_r + \frac{1}{r}\partial_r\right) -(\mathcal{M}^2-E^2) + \frac{2E\alpha}{r} + \frac{\hat{Z}}{r^2},
\end{eqnarray}
where
\begin{eqnarray}
\label{g11}
\hat{Z} = -\frac{(-i\partial_\theta \pm \frac{n}{4})^2}{(1-\frac{n}{6})^2}-\sigma_3 \frac{(-i\partial_\theta \pm \frac{n}{4})}{(1-\frac{n}{6})}+(\alpha^2-\frac{1}{4})-i\alpha\sigma_2.
\end{eqnarray}
The Green's function in the presence of the conical defect can thus be written as 
\begin{eqnarray}
\label{gfcone}
G(\vec{r},\vec{r_0};E)=-i[\hat{\mathcal{H}}+\mathcal{M}]\int_{0}^{\infty}ds e^{irs\{\hat{\mathcal{H}}^2-\mathcal{M}^2\}}\sqrt{\frac{r}{r_0}}\delta(r-r_0)\delta(\theta-\theta_0). 
\end{eqnarray}

Let $|\mathcal{Z}(\theta)>$ be
the normalized eigenfunction of the operator $\hat{Z}$,
\begin{eqnarray}
\label{g12.1}
\hat{Z}|\mathcal{Z}^\kappa (\theta)>=-\kappa^2|\mathcal{Z}^\kappa (\theta)>,
\end{eqnarray}
where $-\kappa^2$ denotes the eigenvalue of $\hat{Z}$. $|\mathcal{Z}(\theta)>$ has 
two components corresponding to the two sublattices of the graphene. Now consider the 
projectors \cite{martin,mil1} $\Omega^\kappa(\theta,\theta_0) = |\mathcal{Z}^\kappa (\theta)><\mathcal{Z}^\kappa (\theta_0)|$ which 
satisfy the relations 
\begin{eqnarray}
\label{g12a}
\hat{Z}\Omega^\kappa(\theta,\theta_0)=-\kappa^2\Omega^\kappa(\theta,\theta_0)
\end{eqnarray}
and
\begin{eqnarray}
\label{g12b}
\sum_\kappa\Omega^\kappa (\theta,\theta_0)= \delta(\theta-\theta_0). 
\end{eqnarray}
 The eigenvalue $-\kappa^2$ is determined from Eq.(\ref{g12a}), which can be written as
\begin{eqnarray} 
\label{g13}
\left( 
\begin{array}{cc}
-\frac{(-i\partial_\theta \pm \frac{n}{4})^2}{(1-\frac{n}{6})^2}-\frac{(-i\partial_\theta \pm \frac{n}{4})}{(1-\frac{n}{6})}+\alpha^2-\frac{1}{4}+\kappa^2 &  -\alpha \\
\alpha & -\frac{(-i\partial_\theta \pm \frac{n}{4})^2}{(1-\frac{n}{6})^2}+\frac{(-i\partial_\theta \pm \frac{n}{4})}{(1-\frac{n}{6})}+\alpha^2-\frac{1}{4}+\kappa^2
\end{array}
\right)\left( 
\begin{array}{cc}
 \Omega^{\kappa}_{11} & \Omega^{\kappa}_{12}\\
\Omega^{\kappa}_{21} & \Omega^{\kappa}_{22}
\end{array}
\right)=0.
\end{eqnarray} 
Using the completeness of the total 
angular momentum eigenfunctions, the components of the projectors 
can be expressed as a linear combination of the functions $e^{ij(\theta-\theta_0)}$ and $e^{-ij(\theta-\theta_0)}$,
where $j$ is taken to be a positive half integer. Using this and solving Eq.(\ref{g13}) we finally get $\kappa = (\eta \pm \frac{1}{2})~$ with $\eta= \sqrt{\nu^2 - \alpha^2}$, $\nu = \frac{j\pm\frac{n}{4}}{1-\frac{n}{6}}$ and  
\begin{eqnarray}
 \label{g14}
\Omega_{11} &=& \frac{(\eta+\nu)}{4\pi\eta}e^{ij(\theta-\theta_0)}+\frac{(\eta-\nu)}{4\pi\eta}e^{-ij(\theta-\theta_0)} = \Omega^*_{22}\nonumber\\
\mbox{and}~~ \Omega_{12} &=& \frac{-i\alpha}{4\pi\eta} [e^{ij(\theta-\theta_0)}+e^{-ij(\theta-\theta_0)}]= \Omega_{21}.
\end{eqnarray}
Note that these projectors are different from what is obtained for planar graphene. This is due to 
the presence of the quantity $\nu$ in the projectors which explicitly depend on the number of 
sectors $n$ removed from the graphene plane to introduce the conical defect. The Green's function 
in the presence of the topological defect can then be written as
\begin{eqnarray}
\label{g14.1}
G(\vec{r},\vec{r_0};E)&=& -i\left[\sigma_3\left(E+\frac{\alpha}{r}\right)-\frac{i\sigma_2}{r(1-\frac{n}{6})}\left(-i\partial_\theta \pm \frac{n}{4}\right) - \sigma_1 \left(\partial_r + \frac{1}{2r}\right)+\mathcal{M}\right]\nonumber\\
&\times& \sum_\kappa\Omega^\kappa (\theta,\theta_0)\int_{0}^{\infty}ds e^{2isE\alpha-2is\left\{\frac{1}{2}\left[-r \partial^{2}_r - \partial_r + \frac{\kappa^2}{r}\right] +(\mathcal{M}^2 - E^2) \frac{r}{2}\right\}}\sqrt{\frac{r}{r_0}}\delta(r-r_0).  
\end{eqnarray}
The Green's function explicitly depends on the conical defect through 
the appearance of the quantity $n$ in it and it is different from the Green's function for the planar case.

In order to proceed we follow the operator technique for Green's function in QED developed in \cite{mil1}, 
which was adapted to the problem of planar graphene in \cite{mil2}. The key point in this approach is to note that the operators
$\frac{1}{2}\left[-r \partial^{2}_r - \partial_r + \frac{\kappa^2}{r}\right]$, $r$ and $-i\left(r\partial_r + \frac{1}{2}\right)$ form the generators of the $O(2,1)$ algebra . Using the commutation relations of these $O(2,1)$ generators, the action of the exponential in Eq.(\ref{g14.1}) on $\delta(r-r_0)$ can be evaluated. The integral representation of the Green's function can be written as 
\begin{figure}
[ht] 
\centering
\includegraphics[bb= 320 14 12 190]{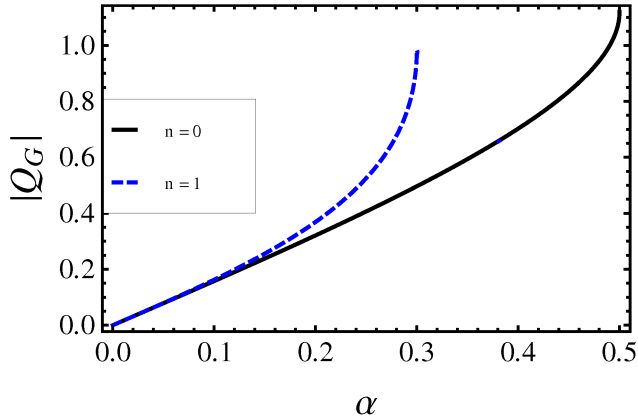}
\caption{Dependence of polarization charge on subcritical Coulomb potential using Green's function.}
\label{fig:1.1}
\end{figure}
\begin{eqnarray}
\label{g15}
G(\vec{r},\vec{r_0};E)&=& i\left[\sigma_3\left(E+\frac{\alpha}{r}\right)-\frac{i\sigma_2}{r(1-\frac{n}{6})}\left(-i\partial_\theta \pm \frac{n}{4}\right) - \sigma_1 \left(\partial_r + \frac{1}{2r}\right)+\mathcal{M}\right]\nonumber\\
&\times& \sum_\kappa\Omega^\kappa (\theta,\theta_0)\int_{0}^{\infty}ds \frac{K}{\sin(Ks)}\mbox{exp}\left[2isE\alpha + iK(r + r_0)\cot(Ks) - i\pi\kappa\right] J_{2\kappa}\left(\frac{2K\sqrt{rr_0}}{\sin(Ks)}\right),
\end{eqnarray}
where $K=\sqrt{(\mathcal{M}^2-E^2)}$. The Green's function in Eq.(\ref{g15}) has been expressed in terms 
of certain operators and an integral. The information about the conical defect appears in the Green's 
function through $n,~\kappa$ and $\Omega^\kappa$. When $n=0$, Eq.(\ref{g15}) reduces to the Green's 
function for planar graphene. The crucial point to note at this stage is that the integral in Eq.(\ref{g15}) 
has the same analytical structure as that for the planar graphene, which is a consequence of our treatment 
of the conical defect. Due to this, we can follow the analysis in \cite{mil2} in order to evaluate the 
integral in Eq.(\ref{g15}). This calculation involves the choice of a suitable contour of integration \cite{mil3,mil2}. 
The resulting expression for the induced charge  is divergent, which requires a renormalization as in the 
case of QED \cite{mil3}. The condition that the total induced charge is zero is imposed as a physical 
requirement in the renormalization process. This is possible due to the presence of the infrared cutoff 
given by the quasiparticle mass $\mathcal{M}$ and it is for this reason that $\mathcal{M}$ was introduced. 
The details of the calculation in the planar case can be found in \cite{mil3,mil2}, following which, 
we get the induced charge density in the $\mathcal{M} \rightarrow 0$ limit as 
\begin{eqnarray}
 \label{g17}
\rho_I(\vec{r}) = Q_G \delta(\vec{r}), 
\end{eqnarray}
where the induced charge $Q_G$ is given by 
\begin{eqnarray}
 \label{g18}
Q_G = -e\left[\frac{\pi \alpha}{2} + \frac{8}{\pi}\sum_j \mbox{Im}\left\{\frac{i\alpha}{2\nu} + \frac{\mbox{ln}(\eta-i\alpha)}{2} + \mbox{ln}\Gamma(\eta-i\alpha)- (\eta - i\alpha)\psi(\eta - i\alpha) - i\alpha\nu\psi^\prime(\nu)\right\}\right],
\end{eqnarray}
where $-e$ denotes the electronic charge and $\psi$ and $\psi^\prime$ denote 
the digamma function and its derivative respectively. Thus, for a subcritical Coulomb charge 
in gapless graphene with a conical defect, the polarization charge is localized at the 
apex of the graphene cone, which is also the position of the external Coulomb charge. 

In order to illustrate the effects of the conical defect, in Fig.(\ref{fig:1.1}) 
we have plotted the dependence of the polarization charge $|Q_G|$ (in units of 
electronic charge) on the Coulomb charge strength $\alpha$. The plot shows the behaviour of the induced charge for 
both $n=0$ (planar graphene) and $n=1$, which corresponds to the conical defect. 
For $n=0$, the plot shows a marked deviation from linear profile at around $\alpha = 0.5$, 
which corresponds to the critical value of the charge in the planar case. 
For $n=1$, the plot shows a similar deviation starting with a smaller value of $\alpha = 0.3$. 
This behaviour is consistent with the fact that the critical charge in the presence of 
the conical defect depends on $n$ and that for $n=1$, the critical charge has a value 
given by $0.3$ \cite{critical1}. To get an analytical feeling for this phenomenon, 
it is useful to consider the expansion of $Q_G$ as a function of $\alpha$. For $n=1$, we get
\begin{eqnarray}
 \label{g19}
Q_G = -e\left[\frac{\pi \alpha}{2} + 5.289 \alpha^3 + 27.437 \alpha^5.....\right].
\end{eqnarray}
For the planar case, the leading term has the same value but the coefficients of  
$\alpha^3$ and $\alpha^5$ have values $0.783$ and $1.398$ respectively \cite{mil2}. 
From these values it is clear that the deviation from linearity sets in at a 
lower value of $\alpha$ in the presence of the conical defect, which is a clear 
indication of the influence of the topological defect on the magnitude of the induced charge.

\subsection{Induced charge from Friedel Sum Rule}

The Friedel sum rule for a gapless graphene cone is given by
\begin{eqnarray}
 \label{f10}
\Delta N = \frac{4}{\pi}\sum_j \left[\delta_j \left(E_F\right)\right],
\end{eqnarray}
where $\Delta N$ represents the change in the number
of charged particles due to the Coulomb potential, 
$\delta_j$ represents the scattering phase shift in the $j$-th 
angular momentum channel and $E_F$ is the Fermi energy \cite{lin3}.  
\begin{figure}
[ht] 
\centering
\includegraphics[bb= 270 14 12 180]{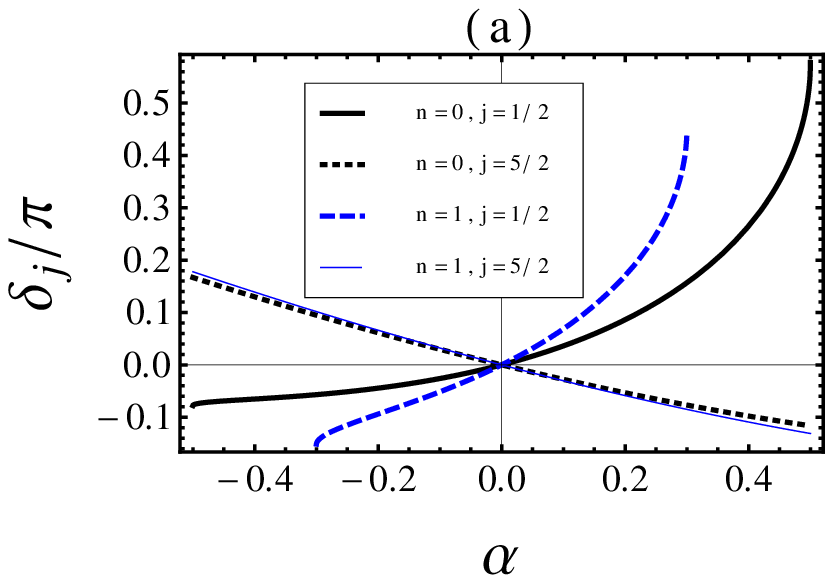}
\includegraphics[bb= -230 14 12 140]{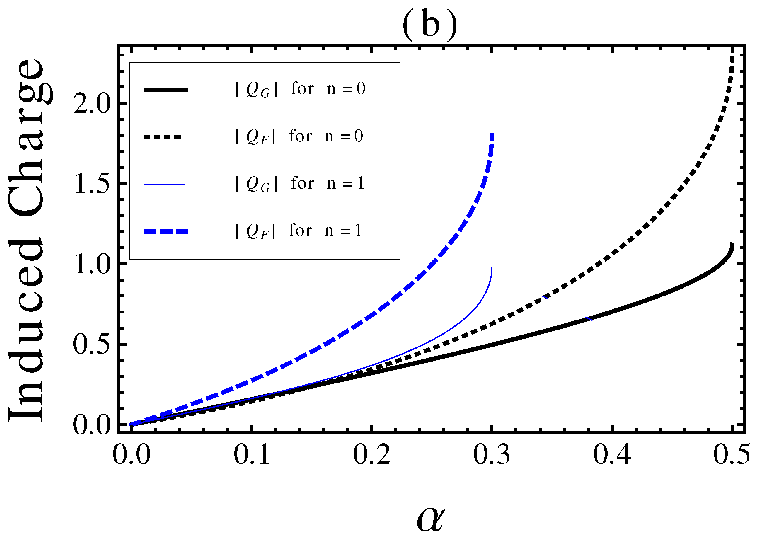}
\caption{(a)Dependence of $\frac{\delta_{j}}{\pi}$ on subcritical Coulomb potential is shown
          for $j=\frac{1}{2}~\mbox{and}~\frac{5}{2}$ when $n=0~\mbox{and}~1$.
         (b)Dependence of polarization charge on subcritical Coulomb potential is shown using both
         Friedel sum rule and Green's function technique. We have considered the effect of $j=\frac{1}{2}$ only.
         The solid lines correspond to Green's function technique and the dashed lines correspond to Friedel sum rule.
         In this paper the positive values of $\alpha$ correspond to an attractive Coulomb potential.}
\label{fig:2}
\end{figure}
For planar gapless graphene, the induced charge from the Friedel sum rule 
was discussed in \cite{levi1} in detail. Here we will discuss the problem in the presence of the conical defect. 
The expression of scattering phase shift for the massless graphene with 
a conical defect in presence of a subcritical Coulomb charge is given by
\begin{eqnarray}
 \label{f17}
\delta_{j}(k)= \alpha \mbox{sgn}(E) \ln(2kr) - \mbox{sgn} (E) \left(\arg[\sqrt{(\eta-i\alpha)}]+\arg[\Gamma(1+\eta+i\alpha)]\right) - \frac{\pi}{2}(\eta-|\nu|),
\end{eqnarray}
where $k=-E$. 
The positive energy values correspond to the electrons whereas negative  energy values  correspond to the holes
in our work. The detail of the calculation can be found in Appendix A. From Eq.(\ref{f17})
we can see that the scattering phase shift contains the term $\nu$ which depends on $n$. 
Note that the scattering phase shift is asymmetric with respect to 
the sign of $\alpha$. For planar graphene, the phase shift has been discussed in \cite{levi1,rmp4,novikov1} 
in detail and we recover the same result when $n=0$.  

The polarization charge is now defined as $Q_F= -e\Delta N$. We have shown the effect of two different angular momentum channels 
$j=\frac{1}{2}~\mbox{and}~\frac{5}{2}$
on the scattering phase shift in Fig.(\ref{fig:2}a). The plot shows that
only in the lowest angular momentum channel $j=\frac{1}{2}$ the sign of
$\delta_{j}$  has an expected nature for attractive and repulsive Coulomb 
potential. This feature has already been noted for planar graphene \cite{rmp4} and continues to hold in the presence of 
the conical defect as well. Due to this reason, we consider the effect of the lowest angular momentum channel only when calculating $|Q_F|$.
In our plot we have neglected the customary $\alpha \mbox{sgn} (E) \ln(2kr)$, which is due to the Coulomb potential \cite{levi1}. 

In Fig.(\ref{fig:2}b) a comparison between $|Q_G|$ and $|Q_F|$ is shown. For $n=0$, both of them have similar behaviour 
for low values of $\alpha$. In the presence of the conical defect, the difference is apparent from $\alpha \approx 0$. 
It is known that there are subtleties associated with the Friedel sum rule in the presence of singularities \cite{graphs}. 
Whether similar effects are playing a role in the presence of a topological defect requires further investigation. 

\begin{figure}
[ht] 
\centering
\includegraphics[bb= 270 14 12 180]{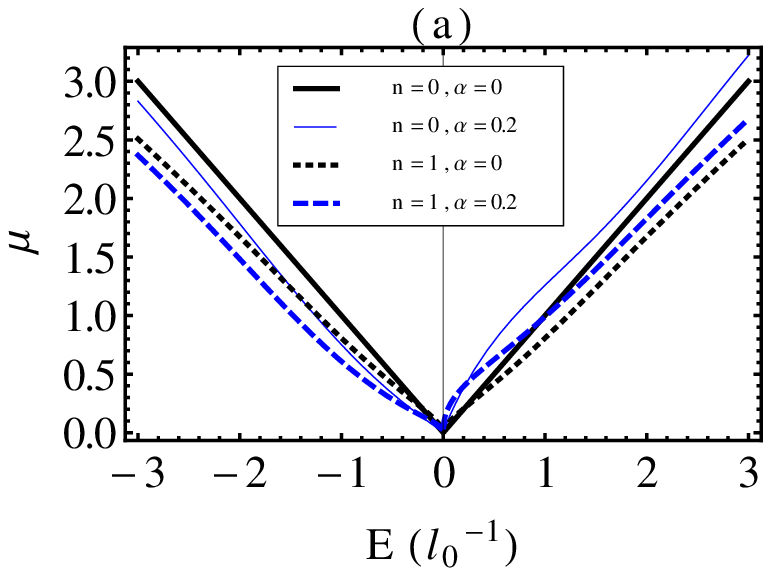}
\includegraphics[bb= -230 14 12 180]{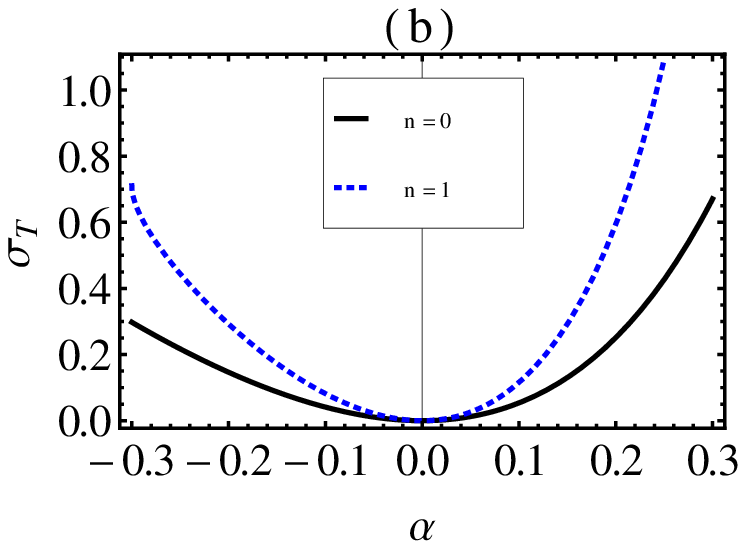}
\caption{(a)Dependence of LDOS on energy of the quasiparticles in the subcritical region.
            We have considered $r=l_0$ in our plot.
         (b)Dependence of $\sigma_T$ on Coulomb potential strength is shown.
}
\label{fig:3}
\end{figure}

We would now like to study how the conical topology affects the physical observables such as the local density of states (LDOS) and the 
conductivity in graphene. The LDOS $\mu$ is defined as
\begin{eqnarray}
 \label{ldos}
\mu = \frac{4}{\pi}\sum_j |\Psi(E,r)|^2 ,
\end{eqnarray}
 where $\Psi(E,r)$ represents the normalized eigenstates 
directly obtained from the solution of the Dirac equation. We have plotted the dependence of $\mu$
on the energy of the quasiparticles in a graphene cone in Fig.(\ref{fig:3}a) where 
the distance from the Coulomb charge has been chosen as $r=l_0$ 
where $l_0$ is of the order of the lattice scale in graphene and energy is given in units of $l_0^{-1}$.
When a positive 
Coulomb charge is introduced in our system it attracts negative charge carriers and repels the holes. A
s a result, the density of negative charge carriers 
increases and that of the positive charge carriers decreases near the impurity.
The same effect can be observed from this plot. In the presence of a conical defect with $n=1$, the plot shows that the density 
of both positive and negative charge carriers decrease compared to the planar case. It may be possible to detect this effect of the conical 
defect with the STM. 

The transport cross section $\sigma_{\mbox{trans}}$ is related to the scattering phase shift by the relation
\begin{eqnarray}
 \label{c1}
\sigma_{\mbox{trans}} = \frac{4}{k}\sum_{j}\sin^2(\delta_j - \delta_{j+1}) = \frac{4}{k}\sigma_T,
\end{eqnarray}
where $k=-E$ and $\delta_j$ is given by Eq.(\ref{f17}) for the subcritical regime. 
The conductivity is inversely proportional to the transport cross-section \cite{ando,nomura,levi2,hwang,novikov1}. 
We have plotted the dependence of $\sigma_T$ on subcritical Coulomb potential strength in Fig.(\ref{fig:3}b). 
The plot is asymmetric with respect to the sign of the potential because the 
scattering cross section of conduction electrons depends on the polarity of the external Coulomb charge.
We can see from the plot that when the value of $n$ increases from $0$ to $1$, the value of $\sigma_T$
also increases. This can be interpreted to be arising from additional interactions associated with 
the conical defect, which leads to a decrease of the conductivity. 

\section{Screening of supercritical Coulomb charge in graphene with conical defect}


For planar graphene, the induced charge density in the supercritical regime has been calculated
using both scattering phase shifts \cite{levi1} as well as the  Green's function \cite{nishida}.
Here we analyze the effect of conical topology on the induced charge in the supercritical regime using the scattering 
phase shift. In the presence of a conical defect, the supercritical regime sets in when $\alpha > \nu$ for given values of $n$ and $j$ \cite{critical1}. 
Defining $\beta = i\eta=\sqrt{\alpha^2 - \nu^2}$ and using zigzag edge boundary condition, the scattering 
phase shift can be written as \cite{critical1} 
\begin{eqnarray}
\label{f22}
\delta_j (k) = \mbox{arg}[e^{i\mbox{sgn}(E)\xi(k)} + c e^{-i\mbox{sgn}(E)\xi(k)}] -\mbox{sgn}(E) \mbox{arg}(h_{\alpha,\beta}) + \alpha \mbox{sgn}(E) \ln(2kr),
\end{eqnarray}
where $k=-E$, $c=e^{\mbox{sgn}(E)\pi \beta}\zeta \frac{h_{\alpha,-\beta}}{h_{\alpha,\beta}}$, 
$h_{\alpha,\beta}=\frac{\Gamma(1+2i\beta)}{\Gamma(1+i\beta-i\alpha)}$,
$\zeta=\sqrt{\frac{\alpha+\beta}{\alpha-\beta}}$ and 
$e^{2i\xi(k)} = \frac{i(1+i\zeta)}{(1-i\zeta)}e^{2i\beta \mbox{ln}(-2\mbox{sgn}(E)kl_0)}$.
\begin{figure}
[ht] 
\centering
\includegraphics[bb= 270 14 12 180]{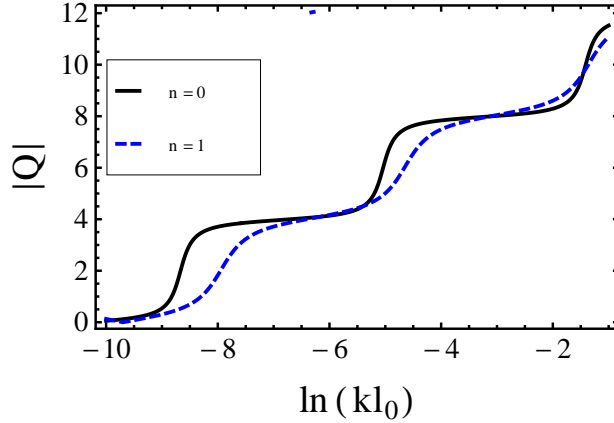}
\caption{
          Using Friedel sum rule dependence of polarization charge on wavenumber $kl_0$ is shown
            in the units of electronic charge.
           Here we have considered the effect of the holes $(\mbox{sgn}(E)=-1)$ for the lowest 
           angular momentum channel $j=\frac{1}{2}$ where $\alpha=1$.}
\label{fig:4}
\end{figure}
The scattering phase shift in the supercritical region has a strong energy dependence
through the first term in the right hand side of Eq.(\ref{f22}). As discussed in \cite{levi1}, 
the induced charge within a radius $r$ is not affected by modes with  
wavelengths larger than $r$. Thus we get
 \begin{eqnarray}
\label{f23}
Q(r)=\frac{-4e}{\pi}\sum_{\nu<\alpha} \delta_{j} (k\sim\frac{1}{r}).
\end{eqnarray}
Neglecting  the customary $\alpha \mbox{sgn}(E) \ln(2kr)$ term of Eq. (\ref{f22}) 
and using Eqns. (\ref{f22}) and (\ref{f23}), $Q(r)$ can be expressed in the units of electronic charge as 
\begin{eqnarray}
 \label{f24}
Q(r)=\frac{4\mbox{sgn}(E)}{\pi}\sum_{\nu<\alpha} \left[\beta \mbox{ln}(\frac{r}{2l_0})+\tan^{-1}\frac{F_1(\alpha,\nu)\cos[f(\alpha,\nu)-2\mbox{sgn}(E)\beta \mbox{ln}\left(\frac{l_0}{r}\right)]}{F_2(\alpha,\nu)+\mbox{sgn}(E)F_1(\alpha,\nu)\sin[f(\alpha,\nu)-2\mbox{sgn}(E)\beta \mbox{ln}\left(\frac{l_0}{r}\right)]} + F_3(\alpha,\nu)\right],
\end{eqnarray}
where $F_1(\alpha,\nu)= e^{\mbox{sgn}(E)\pi\beta} \zeta |\Gamma(1+i\beta-i\alpha)|$, $F_2(\alpha,\nu)=  |\Gamma(1-i\beta-i\alpha)|$, $F_3(\alpha,\nu)= -\tan^{-1}\zeta + \mbox{arg}(h_{\alpha,\beta})$
and $f(\alpha,\nu) = -2\mbox{sgn}(E)\tan^{-1}\zeta + 2\mbox{arg}\Gamma(1-2i\beta) + \mbox{arg}\Gamma(1+i\beta-i\alpha) - \mbox{arg}\Gamma(1-i\beta-i\alpha).$ 
In Fig.(\ref{fig:4}), using the Friedel sum rule, the dependence of $|Q(r)|$ on the wavenumber
$kl_0$ is shown  in the units of electronic charge. We consider the effect of the
holes in our plot for which $\mbox{sgn}(E) = -1$. The kinks appearing in the 
plot correspond to the formation of quasibound states.

From $Q(r)$ we can calculate the induced charge density $\rho_I(r)$ in the supercritical 
region of the graphene cone using the relation \cite{levi1}
\begin{eqnarray}
 \label{f25.1}
\rho_I(r)=\frac{1}{2\pi r}\frac{dQ(r)}{dr} = \frac{\sum_{\nu<\alpha}D_{\nu}(r)}{r^2}.
\end{eqnarray}
The coefficient of the power law tail of induced charge density is given by 
\begin{eqnarray}
 \label{f26}
D_{\nu}(r) = \frac{2\beta}{\pi^2}\left[\mbox{sgn}(E)-2\frac{D_1}{D_2}\right] ,
\end{eqnarray}
where
$$D_1 = \left(\frac{F_2}{F_1}\sec\left[f+2\mbox{sgn}(E)\beta \mbox{ln}\left(\frac{r}{l_0}\right)\right]\tan\left[f+2\mbox{sgn}(E)\beta \mbox{ln}\left(\frac{r}{l_0}\right)\right]+\mbox{sgn}(E)\sec^2\left[f+2\mbox{sgn}(E)\beta \mbox{ln}\left(\frac{r}{l_0}\right)\right]\right)
$$
and
$$D_2 =\left(1+\left\{\frac{F_2}{F_1}\sec\left[f+2\mbox{sgn}(E)\beta \mbox{ln}\left(\frac{r}{l_0}\right)\right]+\mbox{sgn}(E)\tan\left[f+2\mbox{sgn}(E)\beta \mbox{ln}\left(\frac{r}{l_0}\right)\right]\right\}^2\right).
$$
The summation is restricted to those values of $\nu$ such that $\alpha > \nu$, which is needed for supercriticality. 
From Eq.(\ref{f26}) we can see that  $D_{\nu}(r)$ is a log-periodic function of the distance from the Coulomb impurity. 
The period of variation and the average value of $D_{\nu}(r)$ are given by $\frac{\pi}{\beta}$  and $-\frac{2\mbox{sgn}(\alpha E)\beta}{\pi^2}$
respectively. These two quantities depend on the sample topology through the term $\beta$.
For $n=0$, the period of variation of $D_{\nu}(r)$ agrees with the result obtained for planar graphene using the 
Green's function technique \cite{nishida} and the average value of $D_{\nu}(r)$ matches with the constant coefficient
of the power law tail given in \cite{levi1}.

\begin{figure}
[ht] 
\centering
\includegraphics[bb= 270 14 12 180]{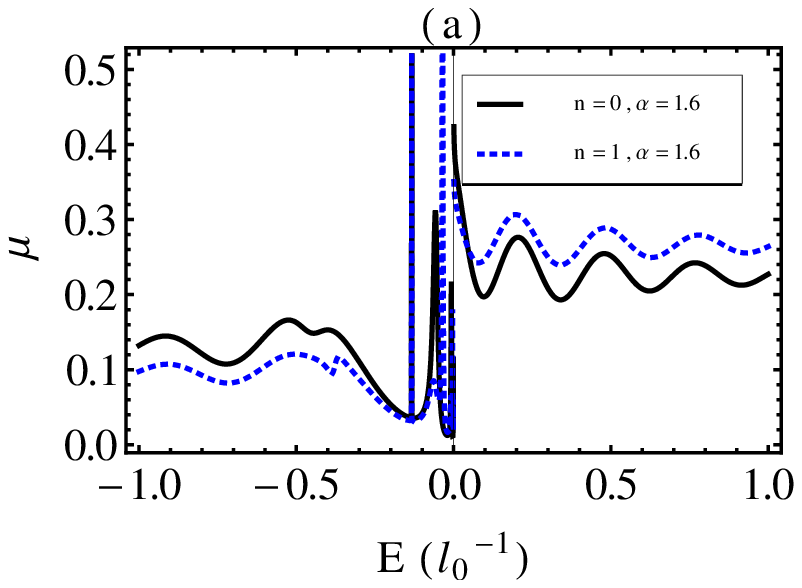}
\includegraphics[bb= -240 4 12 180]{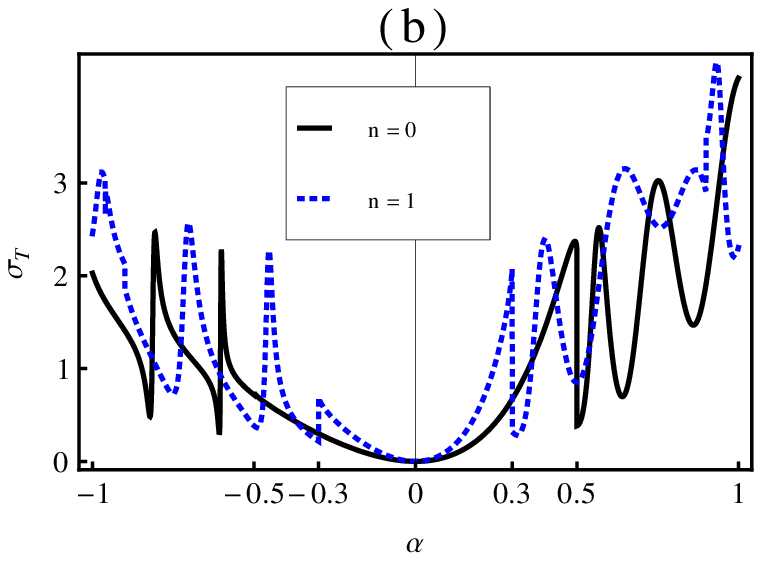}
\caption{(a)Dependence of standing wave oscillations in LDOS on energy of the quasiparticles
         in the supercritical region. We have considered $r=10l_0$ in our plot.
         (b)Dependence of $\sigma_T$ on Coulomb potential strength.
         We have considered $kl_0 = 10^{-5}$ in the plot.}
\label{fig:5}
\end{figure}
The dependence of the LDOS $\mu$ on the energy of the quasiparticles has been plotted in Fig.(\ref{fig:5}a). 
The LDOS has been calculated using eigenstates 
obtained from the solution of the Dirac equation \cite{critical1}, using the angular momentum channels 
for which $\alpha$ is supercritical. The amplitude of the oscillations in LDOS depend on the topological 
defect and they decrease for $n=1$ compared to the planar case.

The transport cross-section of graphene with a conical defect has been calculated using the relation given 
in Eq.(\ref{c1}), where the scattering phase shift $\delta_j$ in the supercritical regime is given 
by Eq.(\ref{f22}). The dependence of $\sigma_T$ on external Coulomb potential strength is shown in 
Fig.(\ref{fig:5}b), for both sub and supercritical regime. The peaks in this plot correspond to the 
quasibound states which occur when the charge is supercritical. It is seen that the peaks start appearing 
for a lower value of $\alpha$ when $n=1$ compared to when $n=0$. This is consistent with the fact that the 
critical charge for $n=1$ has a value $0.3$ compared to the value of $0.5$ for the planar case. As in the subcritical case, the plot for $\sigma_T$ 
is asymmetric with respect to the sign of the potential as the 
transport cross section of conduction electrons depends on the sign of the external Coulomb charge.

\section{Generalized boundary conditions}

The quasiparticles of graphene obey Dirac equation in the long wavelength approximation. 
On the other hand, the external Coulomb charge and the topological defect can interact 
with the graphene lattice and give rise to short range interactions. Such 
short range interactions cannot be directly included in the Dirac equation, which is valid only in the long wavelength limit. 
One way to include the average effect of such short range interactions is through the choice of appropriate boundary 
conditions \cite{critical2}. A technique to classify all such boundary conditions is provided by the 
method of self-adjoint extensions due to von Neumann \cite{reed,falomir,critical2,ksg1}. 
This formalism provides all allowed boundary conditions which are consistent with probability current conservation 
and unitary time evolution. 

The generalized boundary conditions are relevant for graphene with an external charge and a 
conical defect only when $0<\eta<\frac{1}{2}$ \cite{critical3}. In this case, the allowed 
boundary conditions are labelled by a single real parameter $\omega$ defined mod $2 \pi$, also 
known as the self-adjoint extension parameter. The precise value of this parameter cannot be 
predicted by theory alone and has to be determined empirically.
\begin{figure}
[ht] 
\centering
\includegraphics[bb= 270 14 12 180]{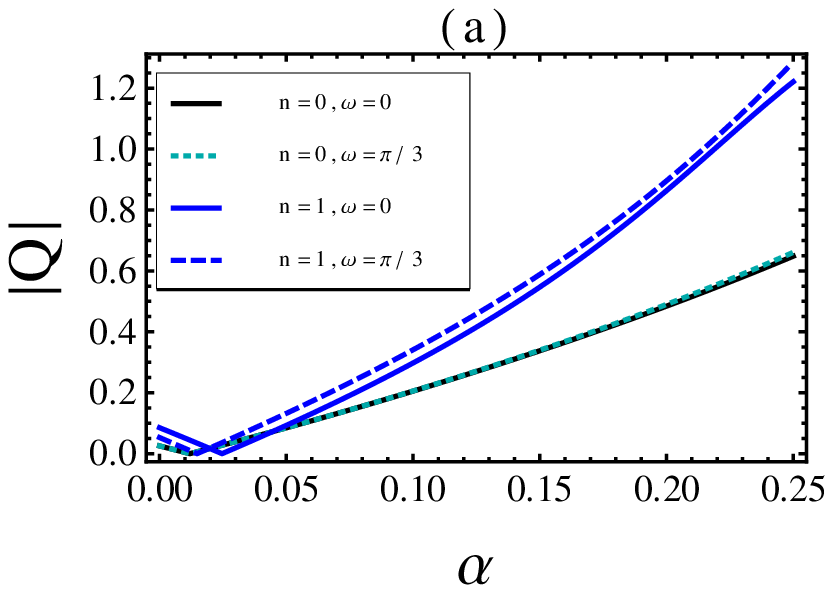}
\includegraphics[bb= -240 14 12 180]{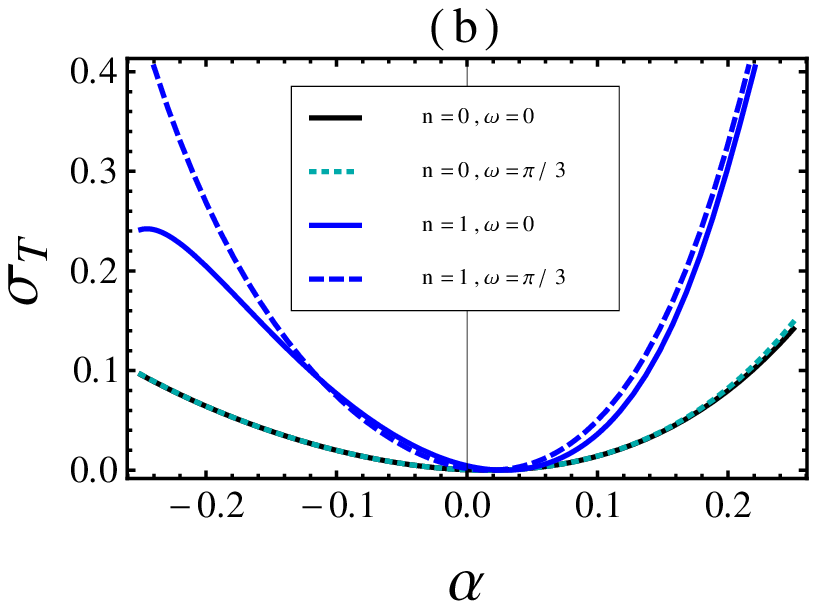}
\caption{(a)Dependence of polarization charge on subcritical Coulomb potential using 
            Friedel sum rule and generalized boundary conditions is shown. The solid lines
            correspond to $\omega = 0$ and the dashed lines correspond to $\omega = \frac{\pi}{3}$.
         (b)Dependence of $\sigma_T$ on Coulomb potential strength is shown
            in presence of generalized boundary conditions. The solid lines
            correspond to $\omega = 0$ and the dashed lines correspond to $\omega = \frac{\pi}{3}$. 
        }
\label{fig:6}
\end{figure}
For this class of generalized boundary conditions, the scattering matrix $S$ 
and the corresponding phase shift $\delta(k)$ are given by
\begin{eqnarray}
\label{sae24}
S=e^{2i\mbox{sgn}(E)\delta(k)}=-\nu e^{2i\mbox{sgn}(E)\alpha \ln(2kr)}\left[\frac{A + B}{\nu^2 A^* + B^*}\right],
\end{eqnarray}
where
\begin{eqnarray}
\label{sae25}
A= \left\{\frac{e^{\frac{i\pi\mbox{sgn}(E)\eta}{2}}-(k)^{2\eta}e^{\frac{-i\pi\mbox{sgn}(E)\eta}{2}}}{\Gamma(1+\eta-i\alpha)\Gamma(1-\eta-i\alpha)}\right\}e^{\frac{i\omega}{2}},
\end{eqnarray}
\begin{eqnarray}
\label{sae26}
B= \left\{\frac{e^{\frac{i\pi\mbox{sgn}(E)\eta}{2}}(-\eta+i\alpha)}{\Gamma(1+\eta-i\alpha)\Gamma(1-\eta+i\alpha)}+\frac{(k)^{2\eta}e^{\frac{-i\pi\mbox{sgn}(E)\eta}{2}}(\eta+i\alpha)}{\Gamma(1-\eta-i\alpha)\Gamma(1+\eta+i\alpha)}\right\}e^{-\frac{i\omega}{2}}.
\end{eqnarray}
See Appendix B for the details. It is clear from Eq.(\ref{sae24}) that the scattering phase shift depends on 
the self-adjoint extension parameter $\omega$ explicitly. 

The induced charge can be calculated from the phase shift using the Friedel sum rule. 
In Fig.(\ref{fig:6}a) we have plotted the induced charge as a function of the Coulomb 
potential strength $\alpha$ with the condition that $0<\eta<\frac{1}{2}$. The polarization charge depends 
on the conical defect as well as on the value of the self-adjoint extension parameter $\omega$. 
The non-zero value of the induced charge for $n=1$ and $\alpha=0$ is indicative of the additional interactions 
due to the topological defect whose average effect is captured by the generalized boundary conditions. 
In Fig.(\ref{fig:6}b) we have also plotted the dependence of $\sigma_T$ on $\alpha$ and $\omega$ for $n=0$ and $1$. 
The contribution of only  $j=\frac{1}{2}$ channel is relevant when $0<\eta<\frac{1}{2}$. It is 
seen that the effect of the generalized boundary condition is more pronounced when $n=1$ 
as compared to the planar case. This is consistent with the fact that generalized boundary 
conditions are more likely to be important in systems with topological defects.

\section{Summary}

In this paper we have analyzed how a conical defect in gapless graphene affects 
the Coulomb screening and related physical observables. In the subcritical regime, 
we have used an operator approach to the Green's function in order to evaluate the 
induced charge. Our calculation shows that the Green's function (\ref{g15}) in the 
presence of the conical defect can be expressed as a product of certain operators 
and an integral, both of which depend on the conical defect through their dependence on $n$. 
The evaluation of the integral in (\ref{g15}) requires a careful choice of the contour and a 
proper renormalization procedure. In our way of treating the conical defect, the integral in (\ref{g15}) 
has the same analytical structure as that for the planar case, although they are not identical. 
This observation allows us to write the result of the integral in (\ref{g15}) from the knowledge 
of the corresponding integral in the planar case, leading to a great simplification of the problem. 
The final result for the induced charge $Q_G$ in (\ref{g18}) explicitly depends on $n$ through $\eta$. 
In other words the induced charge depends on the number of sectors that have been removed from a planar 
graphene lattice in order to form the graphene cone, or equivalently it depends on the sample topology. 
The $n=0$ limit of (\ref{g18}) agrees with the result obtained for planar graphene \cite{mil2}. 

Another interesting effect of the conical topology can be seen from Fig. 2, which shows the 
dependence of the induced charge on the subcritical Coulomb strength $\alpha$. It is clear 
that for $n=1$, the nonlinearity sets in much before that that for $n=0$, the latter being 
the planar case. This onset of nonlinearity happens as the Coulomb strength $\alpha$ approaches 
the critical value. This plot therefore shows that the critical value of the external charge 
actually depends on the topological defect. This striking interplay between dynamics and 
topology can be understood by noting that unlike the planar case, here the critical charge 
is governed not just by the angular momentum $j$ but by the quantity $\nu = \frac{j\pm\frac{n}{4}}{1-\frac{n}{6}}$ \cite{critical1}. 
The behaviour of the induced charge, which is a physical observable, provides a further 
evidence of this interplay. 

In the subcritical regime, we have also calculated the scattering phase shifts and have 
obtained the induced charge using the Friedel sum rule. For the planar case, these 
two approaches give similar values of the induced charge for small values of the 
external Coulomb charge and the results begin to diverge as the critical value is 
approached. In the presence of a topological defect, this divergence sets in for a 
much lower value of $\alpha$. This is possibly related to the subtleties associated 
with the Friedel sum rule in the presence of singularities \cite{graphs}, although 
the exact reason behind this observation requires further investigation. 
For a given value of the subcritical charge, the transport cross-section has a higher 
value in the presence of the conical defect. Similarly, the LDOS in the subcritical 
regime also reflects the effect of the conical defect.

In the supercritical regime, again the induced charge as well as the LDOS and the transport cross-section depend explicitly 
on the conical defect. For the planar case using 
a Green's function approach, it has been observed that the polarization charge density 
in the supercritical regime exhibits a power law tail. The coefficient of the power law 
tail can be written as a sum of functions which vary log periodically with the distance 
from the Coulomb impurity \cite{nishida}. We have obtained the induced charge in this 
regime using the scattering phase shift and the Friedel sum rule. What we find is 
a similar power law tail where the period of the oscillation depends on the sample 
topology through its dependence on $n$. Moreover, the planar limit of this quantity 
as obtained from our calculation agrees with that obtained from the Green's function approach \cite{nishida}. 
The transport cross-section in the supercritical regime shows that the Fano resonances \cite{levi1,levi3}
begin to appear for a lower value of the Coulomb charge in the presence of the conical defect. 
This is consistent with the observation that the critical charge is modified by the conical defect.

Finally, for certain ranges of the system parameters the graphene cone with an external 
Coulomb charge admits generalized boundary conditions \cite{critical2,ksg1,critical3}. 
These boundary conditions incorporate the effect of any short range interactions that 
may have been caused due to the interaction of the external charge as well as the 
topological defect with the graphene lattice. Such interactions cannot be directly 
incorporated in the Dirac equation which is valid only in the long wavelength limit. 
One possible way to take into account such effects is through these generalized 
boundary conditions. We have shown that the observables such as LDOS and the transport cross-section
depends both on the sample topology as well as the choice of the generalized boundary 
conditions.

\appendix
\section{Derivation of scattering phase shift for subcritical region in graphene cone} 

In this appendix the expression of scattering phase shift 
for the massless graphene with a conical defect in presence of a subcritical Coulomb charge
is obtained.
The Dirac equation for the massless graphene cone is given by \cite{critical1} 
\begin{eqnarray} 
\label{A1}
 H \Psi = E \Psi,
\end{eqnarray}
where
\begin{eqnarray} 
\label{A2}
H  = \left( 
\begin{array}{cc}
-\frac{\alpha}{r} &  \partial_r - \frac{i}{r(1-\frac{n}{6})}\partial_\theta \pm \frac{\frac{n}{4}}{r(1-\frac{n}{6})} + \frac{1}{2r}  \\
-\partial_r - \frac{i}{r(1-\frac{n}{6})}\partial_\theta  \pm \frac{\frac{n}{4}}{r(1-\frac{n}{6})}-\frac{1}{2r} & 
 -\frac{\alpha}{r}
\end{array}
\right).\nonumber
\end{eqnarray}
We assume
\begin{eqnarray}
\label{A11}
\Psi(r,\theta)=\sum_j \left( 
\begin{array}{c}
\Psi_{A}^{(j)}(r)\\
i\Psi_{B}^{(j)}(r)
\end{array}
\right)e^{-iEr} {r}^{\eta - (1/2)}e^{ij\theta}.
\end{eqnarray}
For doing the calculations we define two functions $\mathcal{U}^{(j)}(r)$ and $\mathcal{V}^{(j)}(r)$ by
$$\Psi_{A}^{(j)}(r)= \mathcal{V}^{(j)}(r) + \mathcal{U}^{(j)}(r)~~\mbox{and}~~
\Psi_{B}^{(j)}(r)= \mathcal{V}^{(j)}(r) - \mathcal{U}^{(j)}(r)$$.
Using these functions from Eq.(\ref{A1})
we get
\begin{equation} \label{A14}
 x \frac{d^{2} \mathcal{V}^{(j)}(x) }{d x^{2}} + (1 + 2\eta - x)\frac{d \mathcal{V}^{(j)}(x)}{d x} - \left ( \eta + i \alpha \right ) \mathcal{V}^{(j)}(x) = 0,
\end{equation}
where $x= -2ikr$, with  $k=- E$.

The solution of Eq. (\ref{A14}) which obeys regularity at the origin is given by
\begin{eqnarray}
 \label{A15}
\mathcal{V}^{(j)}(x) = C M \left ( \eta + i\alpha,~ 1 + 2\eta,~x \right).
\end{eqnarray}
Here $M(a,b,x)$ is the confluent hypergeometric function \cite{stegun} and $C$ is a constant which 
depends on the energy of the system.

From Eq.(\ref{A15}) and Eq.(\ref{A1}) we have
\begin{eqnarray}
 \label{A16}
\mathcal{U}^{(j)}(x) = C \frac{(\eta+i\alpha)}{\nu} M \left ( 1+\eta + i\alpha,~ 1 + 2\eta,~x \right).
\end{eqnarray}
Using the asymptotic form of $\mathcal{V}^{(j)}(x)$ and $\mathcal{U}^{(j)}(x)$ we can find
out the expression of the scattering phase shift to be
\begin{eqnarray}
 \label{A17}
\delta_{j}(k)= -\mbox{sgn} (E)\left[-\alpha \ln(2kr) + \arg[\Gamma(1+\eta+i\alpha)] - \frac{1}{2}\tan^{-1}\left(\frac{\alpha}{\eta}\right)\right]- \frac{\pi \eta}{2} + \left|\frac{\nu \pi}{2}\right|.
\end{eqnarray}
Eq.(\ref{A17}) is same as Eq.(\ref{f17}). For $n=0$ this result reduces to the one given in \cite{novikov1}.

\section{Derivation of scattering phase shift with generalized boundary conditions}

The radial part of the Dirac operator $H$ in Eq.(\ref{A2}) can be written as 
\begin{eqnarray}
\label{B1}
H_{rad}=
\left( 
\begin{array}{cc}
-\frac{\alpha}{r} & \{ \partial_r +(\nu +\frac{1}{2})\frac{1}{r}\}  \\
-\{\partial_r -(\nu - \frac{1}{2})\frac{1}{r}\} & -\frac{\alpha}{r}
\end{array}
\right).
\end{eqnarray}
The equation which determines the domain of self-adjointness of $H_{rad}$ is given by 
\begin{equation}\label{B2}
 H_{rad}^{\dagger}\Psi_{\pm}=\pm \frac{i}{l}\Psi_{\pm},
\end{equation}
where l is a constant with dimension of length. It can be set equal to unity with a choice of suitable units.
\begin{eqnarray}
\label{B3}
\Psi_{\pm}(r)= \sum_j \bigg (
\begin{array}{c}
\Psi_{A{\pm}}(r)\\
i\Psi_{B{\pm}}(r)
\end{array}
\bigg )
=\sum_j \left( 
\begin{array}{c}
\mathcal{V}_\pm^{(j)}(r) + \mathcal{U}_\pm^{(j)}(r)\\
i(\mathcal{V}_\pm^{(j)}(r) - \mathcal{U}_\pm^{(j)}(r))
\end{array}
\right)e^{\pm \frac{r}{l}} r^{\eta - \frac{1}{2}}.
\end{eqnarray}
The total number of square integrable, linearly independent solutions of Equation(\ref{B2})
gives the deficiency indices $n_{\pm}$ \cite{reed},
which gives a measure of the deviation of the operator $H_{rad}$ from self-adjointness. 
When $n_+ = n_- =0$, $H_{rad}$ is essentially self-adjoint in 
$\mathcal{D}_{0}(H_{rad})$ where $\mathcal{D}_{0} = \mathcal{C}_{0}^{\infty}(\mathcal{R}^{+})$ 
consisting of infinitely differentiable 
functions of compact support in the real half line $\mathcal{R}^{+}$. 
When $n_+ = n_-\neq 0$, $H_{rad}$ is not self-adjoint in $\mathcal{D}_{0}(H_{rad})$ but it 
can admit self-adjoint extensions. When $n_+ \neq n_-$
$H_{rad}$ cannot have self-adjoint extensions \cite{reed}.

The solution for $\Psi_{A+}$ is given by
\begin{eqnarray}\label{B9}
\Psi_{A+}&=& e^{-\frac{r}{l}}\left[U\left(1+\eta-i\alpha,1+2\eta,\frac{2r}{l}\right)-\frac{1}{\nu} U\left(\eta-i\alpha,1+2\eta,\frac{2r}{l}\right)\right]r^{\eta-\frac{1}{2}}.
\end{eqnarray}
As $r\rightarrow\infty,~\Psi_{A+}\rightarrow0$ and when $r\rightarrow0$,
\begin{equation}\label{B10}
\int |\Psi_{A+}|^2 r dr \sim \int r^{-2\eta} dr + \mbox{converging terms}.
\end{equation}
Thus $\Psi_{A+}$ is a square integrable function for 
the range $0<\eta<\frac{1}{2}$. Similarly the entire radial wave function 
can be shown to be  square integrable for the specified range of $\eta$ 
and the deficiency index $n_+ = 1$. Similarly we can show that $n_-=1$ when  $0<\eta<\frac{1}{2}$.
Thus for $0<\eta<\frac{1}{2}$, we get $n_+ = n_- = 1$  and the radial Hamiltonian $H_{rad}$
admits a one parameter family of self-adjoint extension \cite{reed}. The domain of self-adjointness of $H_{rad}$ is given by 
${\mathcal{D}}_{\omega}(H_{rad}) = {\mathcal{D}}_{0}(H_{rad})\oplus  \{e^{i \frac{\omega}{2}} {\Psi}_{+} + e^{-i\frac{\omega}{2}}{\Psi}_{-}\},$
where $\omega\in R ~\mbox{mod} ~2\pi$ is the self-adjoint extension parameter. 

As $r\rightarrow0$,
\begin{eqnarray}\label{B14}
\Psi_{A_+}&=& \frac{\pi}{\nu \sin \pi (1+2\eta)} \left[\frac{(\nu+\eta+i\alpha)}{\Gamma(1-\eta-i\alpha)\Gamma(1+2\eta)}r^{\eta-\frac{1}{2}}-\left(\frac{2}{l}\right)^{-2\eta}\frac{(\nu-\eta+i\alpha)}{\Gamma(1+\eta-i\alpha)\Gamma(1-2\eta)}r^{-\eta-\frac{1}{2}}\right]
\end{eqnarray}
and
\begin{eqnarray}\label{B15}
\Psi_{A_-}&=& \frac{\pi}{\nu \sin \pi (1+2\eta)} \left[\frac{(\nu+\eta+i\alpha)}{\Gamma(-\eta+i\alpha)\Gamma(1+2\eta)}r^{\eta-\frac{1}{2}}-\left(\frac{2}{l}\right)^{-2\eta}\frac{(\nu-\eta+i\alpha)}{\Gamma(\eta+i\alpha)\Gamma(1-2\eta)}r^{-\eta-\frac{1}{2}}\right].
\end{eqnarray}

Note that the physical scattering states  are given by the solution of Eq.(\ref{A1}), from which we get
\begin{eqnarray}
 \label{B18}
 \Psi_A (x)&=& r^{\eta-\frac{1}{2}} e^{ikr} \{\mathcal{C}_1 \frac{(\eta+i\alpha)}{\nu} M \left ( 1+\eta + i\alpha,~ 1 + 2\eta,~x \right)+ \mathcal{C}_2 x^{-2\eta} \frac{(-\eta+i\alpha)}{\nu} M \left(1-\eta + i\alpha,~1 - 2\eta,~x \right)\nonumber\\
           &+& \mathcal{C}_1 M \left ( \eta + i\alpha,~ 1 + 2\eta,~x \right) + \mathcal{C}_2 x^{-2\eta} M \left(-\eta + i\alpha,~1 - 2\eta,~x \right) \}.
\end{eqnarray}
We now match the behaviour of 
the physical wave function with a typical element of ${\mathcal{D}}_{\omega}(H_{rad})$ as $r \rightarrow 0$.
In the limit $r\rightarrow0$ Eq.(\ref{B18}) gives
\begin{eqnarray}
 \label{B19}
\Psi_A (r)=\mathcal{C}_1\frac{\nu+\eta+i\alpha}{\nu} r^{\eta-\frac{1}{2}} + \mathcal{C}_2\frac{(\nu-\eta+i\alpha)}{\nu}(-2ik)^{-2\eta}r^{-\eta-\frac{1}{2}}.
\end{eqnarray}
In the same limit, the upper component of a typical element of the domain 
${\mathcal{D}}_{\omega}(H_{rad})$ is given by 
\begin{eqnarray}
 \label{B20}
 \lim_{r \rightarrow 0} \lambda(e^{\frac{i\omega}{2}}\Psi_{A+} + e^{\frac{-i\omega}{2}}\Psi_{A-}).
\end{eqnarray}
Comparing (\ref{B19}) and (\ref{B20}) and using (\ref{B14}) and (\ref{B15})we get
\begin{eqnarray}
 \label{B21}
\mathcal{C}_1=\frac{\lambda \pi}{\sin\pi(1+2\eta)}&~&\left[\frac{e^{\frac{i\omega}{2}}}{\Gamma(1-\eta-i\alpha)\Gamma(1+2\eta)}+\frac{e^{\frac{-i\omega}{2}}}{\Gamma(-\eta+i\alpha)\Gamma(1+2\eta)} \right]
\end{eqnarray}
and
\begin{eqnarray}
 \label{B22}
\mathcal{C}_2=-(-ikl)^{2\eta}\frac{\lambda \pi}{\sin\pi(1+2\eta)}&~&\left[\frac{e^{\frac{i\omega}{2}}}{\Gamma(1+\eta-i\alpha)\Gamma(1-2\eta)} + \frac{e^{\frac{-i\omega}{2}}}{\Gamma(\eta+i\alpha)\Gamma(1-2\eta)} \right].
\end{eqnarray}
When $r\rightarrow\infty$ we note that
\begin{eqnarray}
 \label{B23}
\Psi_A (r)&=&(-2ik)^{-\eta} (-i)^{i\alpha}\left[\mathcal{C}_1 \frac{\eta+i\alpha}{\nu} \frac{\Gamma(1+2\eta)}{\Gamma(1+\eta+i\alpha)} + \mathcal{C}_2 \frac{-\eta+i\alpha}{\nu}\frac{\Gamma(1-2\eta)}{\Gamma(1-\eta+i\alpha)}\right]\frac{e^{-i[kr-\alpha \ln(2kr)]}}{\sqrt{r}}\nonumber\\
&+&(-2ik)^{-\eta} (-i)^{-i\alpha}\left[\mathcal{C}_1\frac{\Gamma(1+2\eta)}{\Gamma(1+\eta-i\alpha)}e^{-i\pi(\eta+i\alpha)} + \mathcal{C}_2 \frac{\Gamma(1-2\eta)}{\Gamma(1-\eta-i\alpha)}e^{-i\pi(-\eta+i\alpha)}\right]\frac{e^{i[kr-\alpha \ln(2kr)]}}{\sqrt{r}}. 
\end{eqnarray}
Substituting the expressions of $\mathcal{C}_1$ and $\mathcal{C}_2$ from 
Eq.(\ref{B21}) and Eq.(\ref{B22}) in Eq.(\ref{B23}) the scattering matrix 
$S$ and the corresponding phase shift $\delta(k)$ are obtained as
\begin{eqnarray}
\label{B24}
S=e^{2i\mbox{sgn}(E)\delta(k)}=-\nu e^{2i\mbox{sgn}(E)\alpha \ln(2kr)}\left[\frac{A + B}{\nu^2 A^* + B^*}\right],
\end{eqnarray}
where
\begin{eqnarray}
\label{B25}
A= \left\{\frac{e^{\frac{i\pi\mbox{sgn}(E)\eta}{2}}-(kl)^{2\eta}e^{\frac{-i\pi\mbox{sgn}(E)\eta}{2}}}{\Gamma(1+\eta-i\alpha)\Gamma(1-\eta-i\alpha)}\right\}e^{\frac{i\omega}{2}},
\end{eqnarray}
\begin{eqnarray}
\label{B26}
B= \left\{\frac{e^{\frac{i\pi\mbox{sgn}(E)\eta}{2}}(-\eta+i\alpha)}{\Gamma(1+\eta-i\alpha)\Gamma(1-\eta+i\alpha)}+\frac{(kl)^{2\eta}e^{\frac{-i\pi\mbox{sgn}(E)\eta}{2}}(\eta+i\alpha)}{\Gamma(1-\eta-i\alpha)\Gamma(1+\eta+i\alpha)}\right\}e^{-\frac{i\omega}{2}}.
\end{eqnarray}
It is clear from Eq.(\ref{B24}) the scattering matrix and the phase shift 
explicitly depend on the self-adjoint extension parameter $\omega$. 
Eq.(\ref{B24}), (\ref{B25}) and (\ref{B26}) are same as Eq.(\ref{sae24}), (\ref{sae25})
and (\ref{sae26}) given in section $4$, with the choice of the dimensionful parameter $l =1$.

\end{document}